\documentclass[doublecol]{epl2}

\usepackage{graphicx}
\usepackage{dcolumn}
\usepackage{bm}

\title{Simple Model for Wet Granular Materials with Liquid Clusters}
\shorttitle{Simple Model for Wet Granular Materials with Liquid
Clusters}

\author{Namiko Mitarai\inst{1} \and Hiizu Nakanishi\inst{2}}
\shortauthor{N. Mitarai and H. Nakanishi}
\institute{
 \inst{1} Niels Bohr Institute,
Blegdamsvej 17, DK-2100, Copenhagen, Denmark.\\
 \inst{2} Department of Physics, Kyushu University 33, 812-8581, Fukuoka, Japan.
}

\pacs{45.70.-n}{Granular systems}
\pacs{82.70.-y}{Disperse systems; complex fluids}
\pacs{83.80.Fg}{Granular solids}

\abstract{
We propose a simple phenomenological model for wet granular media
to take into account many particle interaction 
through liquid in the funicular state 
as well as two-body cohesive force by a liquid bridge 
in the pendular state.
In the wet granular media with small liquid content,
liquid forms a bridge at 
each contact point, which induces two-body cohesive force
due to the surface tension. As the liquid content increases, 
some liquid bridges merge, and more than 
two grains interact through a single liquid cluster.
In our model, the cohesive force acts between the 
grains connected by a liquid-gas interface.
As the liquid content increases, 
the number of grains that interact through the liquid 
increases, but the liquid-gas interface may decrease
when liquid clusters are formed. 
Due to this competition, our model shows
that the shear stress has a maximum as
a function of the liquid-content.
}

\begin{document}
\maketitle
\section{Introduction}
\begin{figure}[hpt]
\includegraphics[width=0.48\textwidth]{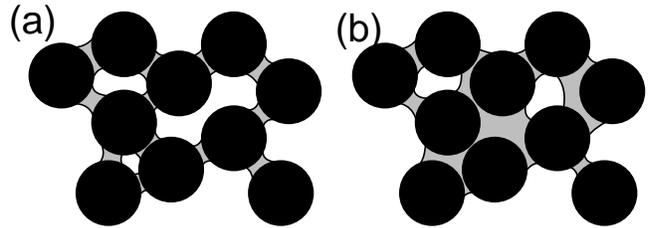}
\caption{Schematic description of the 
distribution of liquids in the wet granular materials.
(a) The pendular state, where 
liquid bridges are formed at contact points.
(b) The funicular state,
where some liquid clusters that connect more than two grains
are formed. }
\label{pendularfunicular}
\end{figure}

It is well known that just adding small amount of liquid
to dry granular materials changes their behaviors
drastically \cite{Mitarai06,Herminghaus05,GranulationReview}.
Liquid forms bridges between grains, 
and the bridge induces cohesion between grains
due to the surface tension. 
When the amount of liquid is small, 
liquid bridges are formed between two grains 
at contact points;
such a state is called  
{\it the pendular state} (Fig.~\ref{pendularfunicular}(a)).
As the amount of liquid is increased,
several liquid bridges are connected to 
form a liquid cluster that includes
more than two grains as in Fig.~\ref{pendularfunicular}(b),
which is called {\it the funicular state}.
As we add the liquid further, 
all the pores will be filled with the liquid.
It is called {\it the capillary state} if 
the liquid pressure is still lower than the air pressure,
and finally the system becomes {\it the slurry state} if 
the amount of liquid is enough 
to make the liquid pressure as high as the air pressure.

The liquid content dependence of the mechanical properties of 
wet granular materials has not been well understood yet.
The most studied case is the pendular state,
both experimentally and numerically
\cite{Mitarai06,Herminghaus05,GranulationReview,BGLL05,RRWNC06,RRNC08,RYR06,RRY06,XOK07,RRNC08}.
There, the situation is rather simple because the grain-grain
interactions, including the cohesion due to liquid bridges,
are mainly two-body interactions. 
However, as the liquid content is increased, 
liquid clusters become non-negligible,
which induce many-body interaction via the liquid.
Methods and/or models to deal with such a situation
have not been established yet and still under active research
(\cite{Fournier05,LWT07,MB07, GLS08-1,GLS08-2,SSBMSBH08}).

Recent experiments show that,
upon increasing the liquid content, there is a maximum
in the material yield stress
at rather low liquid content, in the 
pendular state or the funicular state \cite{MB07,LWT07, Fournier05}.
Especially, 
M\o ller et al. \cite{MB07} showed a clear increase of the shear
modulus with the liquid content in the pendular state
and a drop in the funicular state.

In this paper, we propose a simple 
phenomenological model that takes into 
account the many-body interaction through 
the liquid clusters in the funicular state.
In our model, the cohesive force acts among the 
grains connected by the liquid-gas interface.
As the liquid content increases, 
the number of grains that interact through the liquid 
increases, but the liquid-gas interface may decrease
when the liquid cluster is formed. 
We show that this competition results in a maximum in the shear stress
as a function of liquid-content, which 
may correspond qualitatively with the experimentally 
observed behavior of the shear modulus 
as a function of liquid content.
\section{Model}
In this section, we first introduce a model 
for the pendular state, where the two-body cohesion due to 
liquid bridges is incorporated in 
the soft sphere model of dry grains with 
linear elastic force and dissipation 
(Discrete Element Method, DEM \cite{CS79}).
For the cohesive interaction, history-dependent formation 
of liquid bridge proposed by Schultz {\it et al.} \cite{Schulz03} 
is adopted.
We then extend the model to the funicular state, 
where some pores are filled with
liquid and the many-body interactions 
acts among particles.

\subsection{Model P for the pendular state with 
two-body interaction}
For simplicity, we consider a frictionless
two-dimensional system,
where grains are modeled by polydisperse disks.
Let us consider the interaction between particles $i$ and $j$
of diameters $\sigma_i$ and $\sigma_j$ 
and masses $m_i$ and $m_j$, 
at positions $\bm r_i$ and $\bm r_j$ 
with velocities $\bm v_i$ and $\bm v_j$,
respectively. 
There is no interaction 
between them 
before they collide.
Once they are in contact, i.e., 
$|\bm{r}_{ij}|=|\bm{r}_i-\bm{r}_j|$ becomes less than $(\sigma_i+\sigma_j)/2$,
the force on the particle $i$ by $j$ is given 
as a function of $\Delta\equiv (\sigma_i+\sigma_j)/2-|\bm{r}_{ij}|$
by
\begin{equation}
\bm{F}_{ij}=
f(\Delta)\bm{n}_{ij}
-\eta(\Delta) \bm{v}_{ij}\label{linearDEM}
\end{equation}
with the normal unit vector $\bm{n}_{ij}=\bm{r}_{ij}/|\bm{r}_{ij}|$ 
and the relative velocity $\bm{v}_{ij}=\bm{v}_i-\bm{v}_j$, 
as long as $|\bm r_{ij}|<\alpha(\sigma_i+\sigma_j)/2$;
The parameter $\alpha (\ge 1)$ determines the distance
between grains at which the liquid bridge breaks, and 
once $|\bm r_{ij}|$ exceeds $\alpha(\sigma_i+\sigma_j)/2$,
$\bm{F}_{ij}=0$ and the pair $i, j$ do not interact 
until they are in contact again.

The functions $f(\Delta)$ and $\eta(\Delta)$
in the interaction of Eq.(\ref{linearDEM}) are given by
\begin{equation}
f(\Delta)=\max\left(k \Delta, -F_0\right)
\label{forcelaw}
\end{equation}
and
\begin{eqnarray}
\eta(\Delta)=
\left\{
\begin{array}{ll}
\eta_{\rm{grain}} &\mbox{for $\Delta >0$},\\
\eta_{\rm{liquid}} &\mbox{for $\Delta <0$}.
\label{dissipationcoef}
\end{array}
\right.
\end{eqnarray}
We can interpret that 
$\Delta$ characterizes the overlap (for $\Delta>0$)
or the spacing (for $\Delta<0$) between the grains.
The parameter $k$ characterizes the elastic constant, $\eta(\Delta)$ is the 
damping parameter, and a positive constant
$F_0$ characterizes the cohesion due to a liquid bridge
\footnote{
One might presume that the cohesive force should act even when grains
are in contact.  This can be dealt with, for the present frictionless
system, simply by reinterpreting  $\Delta$ as being measured from the
overlap position where the cohesive force and repulsive force due to
grain elasticity balance.  The dissipation term of eq.(3) should be
modified accordingly, but such a minute adjustment barely affects the
overall system behavior.}.

The damping parameter $\eta(\Delta)$ in eq.(\ref{dissipationcoef})
depends on $\Delta$ because the dissipation when 
grains are in contact (characterized by 
$\eta_{\rm grain}$) and the dissipation via 
liquid viscosity (characterized by $\eta_{\rm liquid}$)
has different physical origin and 
their values can be quite different.

Notice that the model with $\alpha=1$ corresponds to
dry granular materials 
with neither cohesive force nor hysteresis,
and $\alpha$ is greater than unity
for wet cohesive grains.
When $\alpha>1$, there is a hysteresis 
in the interaction \cite{Schulz03}, and this hysteresis
causes dissipation in addition to that caused by
the viscous force.

In this model for the pendular state (here we call it as Model P), 
all interactions are given as two-body forces.
The situation is schematically 
illustrated in Fig.~\ref{liquidbridges}(a);
the attractive force between the particles 
connected by solid arrows represents the 
cohesion due to liquid bridges.

\subsection{Model F for the funicular state with 
many-body interaction}
\begin{figure}[t]
\includegraphics[width=0.23\textwidth]{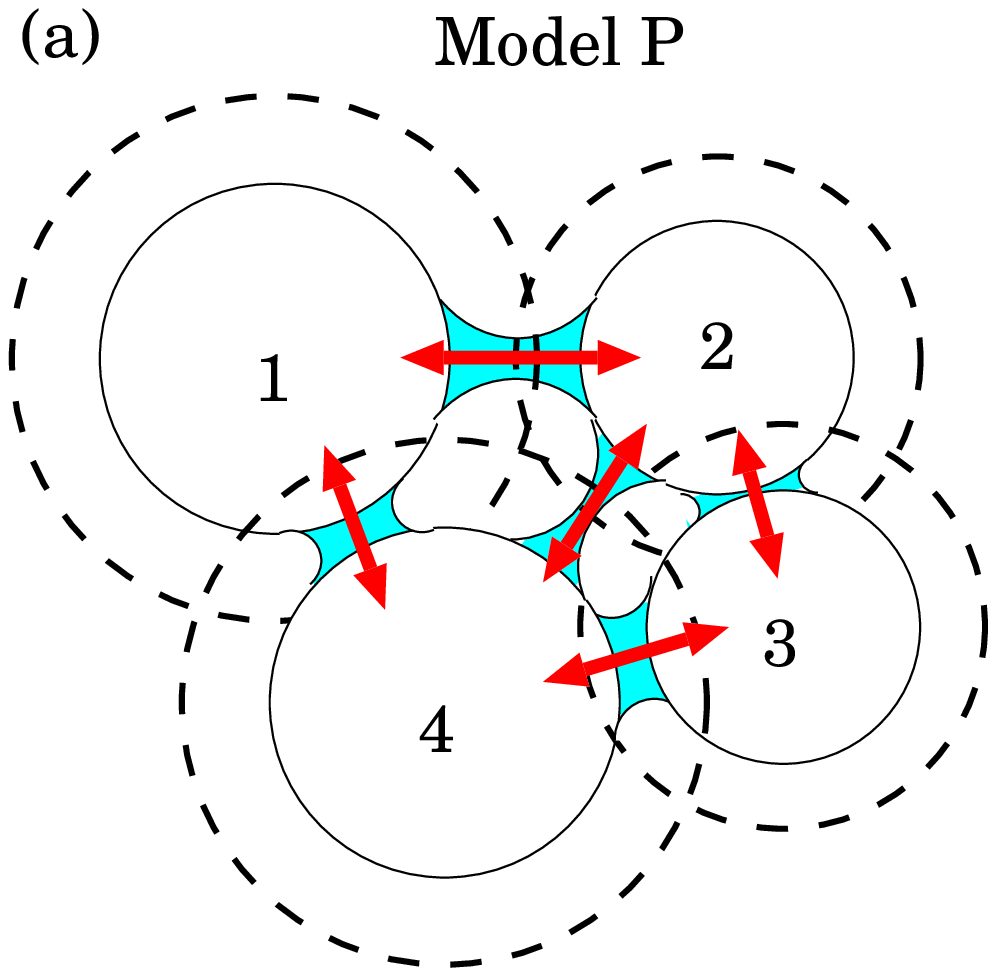}
\includegraphics[width=0.23\textwidth]{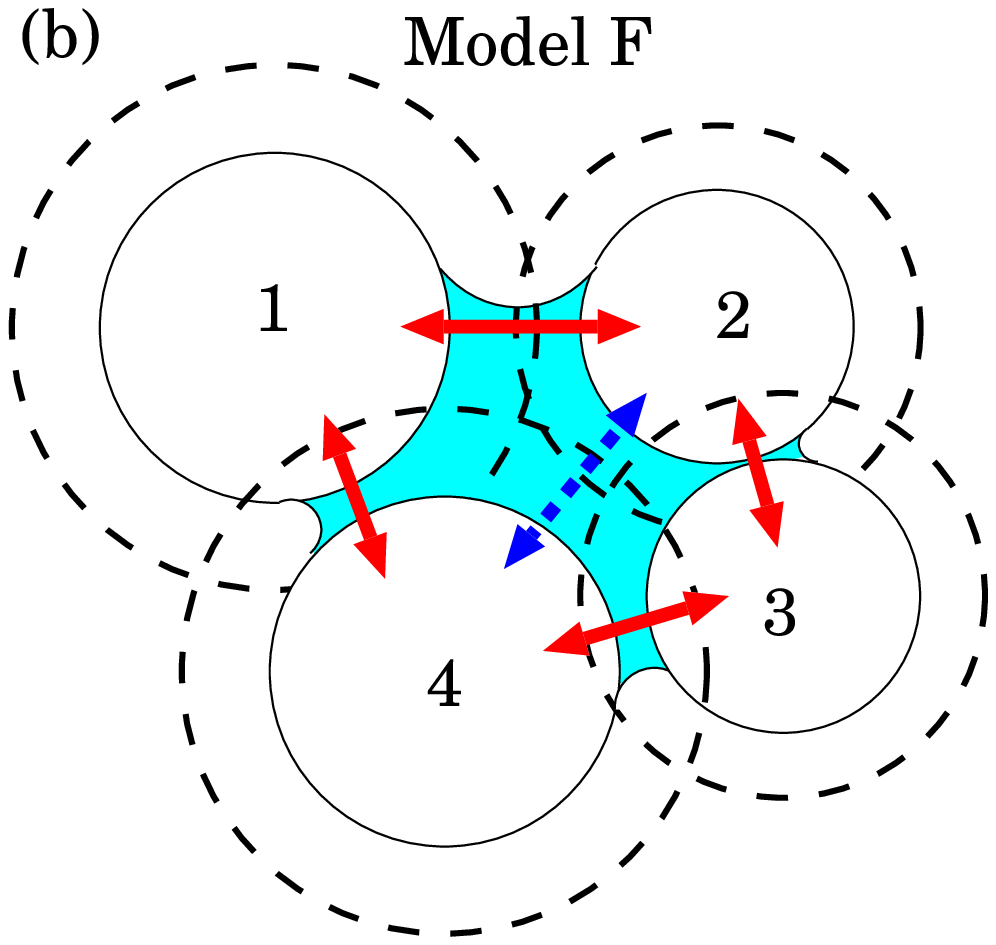}
\caption{Schematic description of interaction 
of (a) the model for the pendular state (Model P) and (b) the model
for the funicular state (Model F).
The solid circles show grains 
and the dotted line indicates 
the interaction range after the formation of 
a liquid bridge.
The solid arrows indicates the pair of grains that 
cohesive force acts.
The dotted arrow in (b) shows ``inactive bonds'',
i.e., they are interacting via liquid cluster 
but the bond exert no cohesive force.
}
\label{liquidbridges}
\end{figure}
When the amount of liquid is increased 
enough for the system to be in
 the funicular state,
some pores are filled by liquid, 
and more than two grains are 
connected by a single liquid-filled region.
Let us consider the situation shown in 
Fig.~\ref{liquidbridges}(b),
where four particles are connected 
by a liquid cluster. 
In this case, there is no 
liquid-gas interface that connects particles 
2 and 4; in contrast to the pendular state
(Fig.~\ref{liquidbridges}(a)),
there is no direct cohesive 
force between them, though 
the cohesive force between the particles 
connected by solid arrows in Fig.~\ref{liquidbridges}(b)
still tend to hold all the particles together.
There should be still some effective cohesive force
between the particles 2 and 4 caused by the 
lower pressure in the liquid phase
than that in the air phase, 
but here we disregard this.

We model such a multi-particle effect by
making the bond inactive when the bond connects the two particles
between which a liquid cluster exists.
The presence of the liquid cluster between
two grains is determined by the simple criterion:
{\it The grains $i$ and $j$ are connected by a liquid cluster
without liquid-gas interface if they are connected by a liquid 
bridge and the number of grains $N_{i,j}$
that are connected to both $i$ and $j$
by liquid bridges is larger or equal to a certain threshold value
$n_{th}$.} The liquid bridge is the same as defined 
in the previous subsection, and we take 
$n_{th}=2$ for the two-dimensional system.
If two grains are connected by a big cluster
without liquid-gas interface, 
the liquid bridge between them is inactive and no 
cohesive force acts through.
An example in two-dimensional system with $n_{th}=2$
is given in Fig.~\ref{liquidbridges}(b).

With this criterion, the force on grain $i$ by $j$
is given not by Eq.~(\ref{linearDEM}) but by 
\begin{eqnarray}
\bm {F}_{ji}&=& f(\Delta)\left[
\Theta(f(\Delta))+
(1-\Theta(f(\Delta)))
\Theta(n_{th}-N_{i,j})\right] \bm n_{ij}\nonumber\\
&&-\eta(\Delta) \bm v_{ij},
\end{eqnarray}
where $f(\Delta)$ is given by Eq.~(\ref{forcelaw})
and $\Theta(x)$ is the Heaviside step function
defined as $\Theta(x)=1$ for $x> 0$ and 
$\Theta(x)=0$ for $x\le 0$.
We call this model for the funicular state as Model F here.
Note that it is straightforward to extend this model 
to a three-dimensional system by choosing 
$n_{th}=3$. 

\begin{figure}[tp]
\centerline{\includegraphics[angle=-90,width=0.4\textwidth]{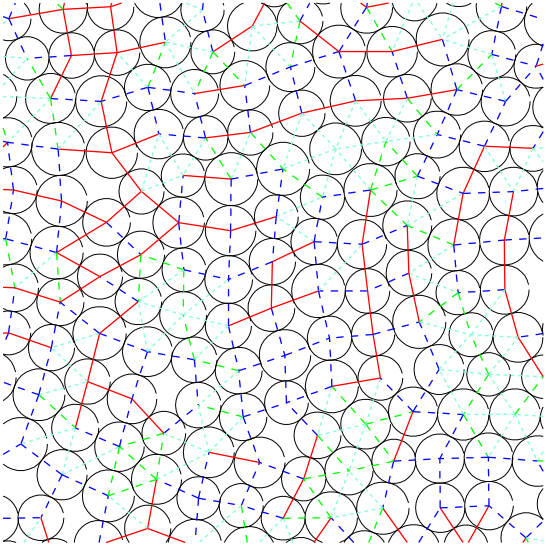}}
\caption{Snapshot of grain-grain interactions.
The interacting grains are connected by bonds.
The cohesive bonds, or the bonds with 
$N_{i,j}<n_{th}$ and $\Delta<0$,
are shown by red solid lines, the repulsive bonds
with liquid interfaces,
or $N_{i,j}<n_{th}$ and $\Delta>0$,
are by blue long-dashed lines,
the repulsive bonds
in liquid clusters,
or $N_{i,j}\ge n_{th}$ and $\Delta>0$,
are by green long-dashed lines,
and the inactive bonds, or
$N_{i,j}\ge n_{th}$ and $\Delta<0$,
by a light blue short-dashed line.
}
\label{network}
\end{figure}

Figure \ref{network} illustrates an example of 
the interaction network in Model F.
The interacting grains are connected by bonds.
The cohesive bonds, or the bonds with 
$N_{i,j}<n_{th}$ and $\Delta<0$,
are shown by red solid lines, the repulsive bonds
with liquid interfaces,
or $N_{i,j}<n_{th}$ and $\Delta>0$,
are by blue long-dashed lines,
the repulsive bonds
in liquid clusters,
or $N_{i,j}\ge n_{th}$ and $\Delta>0$,
are by green long-dashed lines,
and the inactive bonds, or
$N_{i,j}\ge n_{th}$ and $\Delta<0$,
by a light blue short-dashed line.
We can see several clusters of particles connected by 
bonds with $N_{i,j}\ge n_{th}$, which corresponds to liquid clusters.

In this model, the parameter $\alpha$
for the interaction range represents the liquid content.
In Model P, increasing $\alpha$
just increases the number of cohesive bonds.
In Model F, on the other hand, 
not only the number of interacting particles increases,
but also the fraction of bonds that are surrounded by 
other bonds ($N_{i,j}\ge n_{th}$) increases; 
such bonds exert no cohesion. 
Thus, the larger $\alpha$ corresponds to the 
case with more liquid content.

We now examine how the competition between these 
effects of increasing $\alpha$ in Model F
affects the response of the wet granular 
assembly against shear.

\section{Simulation Setup}
\begin{figure}[tp]
\includegraphics[width=0.4\textwidth]{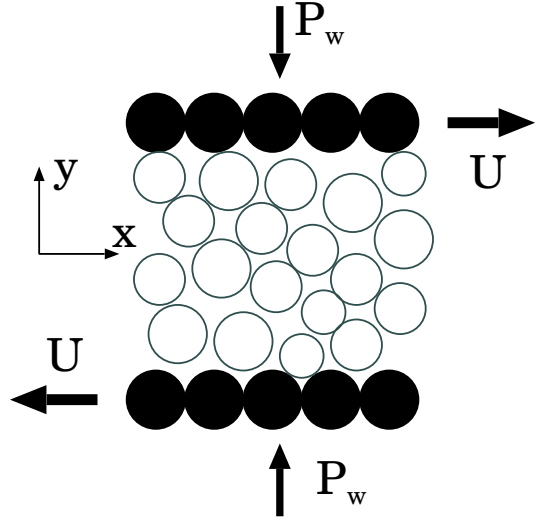}
\caption{
Schematic description of the simulated configuration.
}
\label{system}
\end{figure}
We mainly focus on Model F under shear,  
but we also present the results of Model P for comparison. 

The grains are modeled by polydisperse 
disks whose diameters are uniformly distributed 
between $0.8\sigma$ to $\sigma$,
and all the grains have the same mass $m$.
The system is periodic in the $x$ direction.
In the $y$ direction, there are two parallel 
rough walls of mass $M$, each of which consists of $N_w$ particles
with diameter $\sigma$
glued without spacing (Fig. \ref{system};
the total length of the wall $L=\sigma N_w$). 

The top (bottom) wall is moving 
in the $x$ direction at a velocity $U$ 
($-U$) in order to realize a shear flow with 
velocity gradient in the $y$ direction (Fig. \ref{system}).
A constant pressure $P_w$ is applied 
to the walls in the $y$ direction,
and the $y$ coordinate of the 
top (bottom) wall $Y_t$ $(Y_b)$
obeys the equation of motion 
$M \ddot Y_t = F_{w,t}(t)-P_wL$
($M \ddot Y_b = F_{w,b}(t)+P_wL$),
where $F_{w,t}(t)$ ($F_{w,b}(t)$)
is the $y$-component of the force exerted from 
particles on the upper (bottom) wall at time $t$.

For Model F, we also need to decide whether there is liquid 
bridges among grains that consist the wall;
That affects the formation of the liquid cluster near 
the wall. We initially impose that a 
neighboring pair of grains on the wall
has a liquid bridge with the probability $1/2$,
and the bridges among wall grains do not disappear or newly created 
during the simulation.

Initial configurations for simulations are 
prepared as follows;
$N$ particles are 
distributed randomly with large enough spacing, 
and the wall pressure $P_w$ is applied without shear ($U=0$). 
After waiting long enough so that all the grains are at rest, 
we start to shear the assembly at a constant $U$.

The data are collected in the steady states,
where both the average kinetic energy of 
grains and the 
the $y$-coordinates of the walls become almost 
constant.

The parameters are given in the  
dimensionless form with the unit length
$\sigma$, unit mass $m$, and 
the unit time scale $\sqrt{100m\sigma^2/F_{0}}$
in the following.
In this unit, we choose 
the elastic constant $k$ to be $5\times 10^4$,
the viscous coefficient due to grain contact $\eta_{\rm grain}$
to be $10$
(which gives the restitution coefficient $e_n$ for 
dry grains ($\alpha$=1) around $0.9$), and 
the viscous coefficient due to liquid bridge $\eta_{\rm liquid}$
to be $1$. 
In this paper, we focus on the situation where 
the relative velocity of particles is $O(1)$, thus 
the viscous force is about 1/100 of the cohesive force.
Also, note that the particle deformation is 
typically less than 1\% of its diameter under the force 
applied in the simulations.

We use the wall mass $M=N_w m$, 
and several values of the wall pressure $P_w$.
The system size are mostly $L=40$ and the number of particles 
in the system $N=1600$ unless otherwise noted. 
In most of the cases, the wall velocity $U$ is chosen 
to give the shear rate $\dot \gamma \approx 1$
in the middle of the system, and smaller/larger $U$ are 
also examined to see the tendency.
The liquid-content dependence is studied by changing
the interaction range $\alpha$.
The second-order Adams-Bashforth 
method and the trapezoidal rule are used 
to integrate the equations 
for the velocity and position, respectively,
with a time step for integration $dt=10^{-4}$.

\section{Simulation Results}
\begin{figure}[t]
\includegraphics[width=0.45\textwidth]{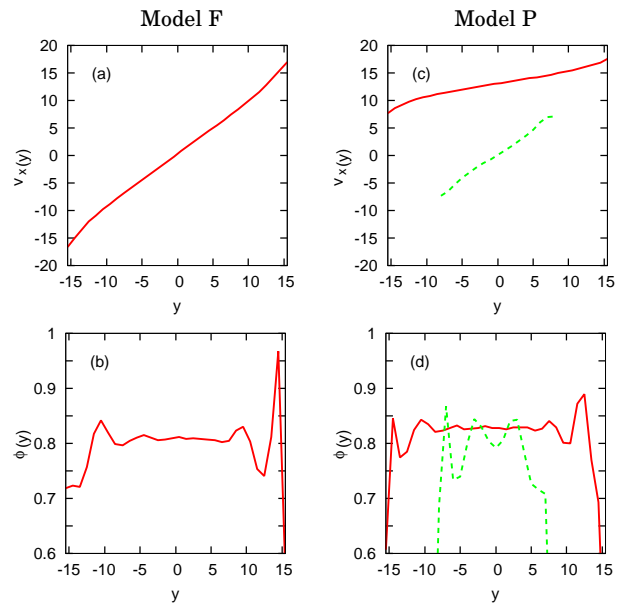}
\caption{
(a) and (b): The velocity (a) and the packing fraction (b) 
profile of the system with $U=18$ for Model F
with $\alpha=1.3$ and $P_w=100$.
In (a), we can see that the shear rate $\dot \gamma$ is almost constant
over the system with $\dot \gamma\approx 1$, 
while a spatial structure can be seen  in the packing fraction profile
near the boundary.
(c) and (d): The velocity (c) and the packing fraction (d) 
profile of the system for Model P
with $\alpha=1.3$ and $P_w=100$.
The data for $L=40$ and $U=18$ is shown by solid lines,
and the data for $L=20$ and $U=9$ is shown by dashed lines.
\label{profile}
}
\end{figure}
We will confirm that the system can achieve
plain shear flows first, 
and then we will investigate
the effect of liquid content on the model
behavior by changing 
the interaction range $\alpha$.

\begin{figure}[t]
\includegraphics[width=0.4\textwidth]{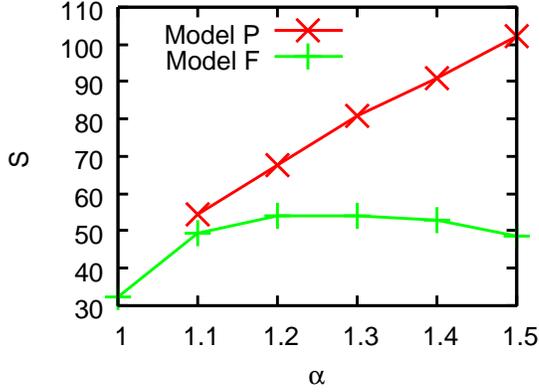}
\caption{
The interaction range $\alpha$ dependence of the 
shear stress $S$ in the case of $P_w=100$ for Model F and Model P
with $\dot \gamma\approx 1$.
(for Model F, $U=18$ is used, while 
for Model P the small system with $L=20$ and
$N=400$ with $U=9$ is studied to avoid the non-uniform shear).
For $\alpha=1$, both models give the same results.
}
\label{shearstress}
\end{figure}
\begin{figure}[t]
\includegraphics[angle=-90,width=0.48\textwidth]{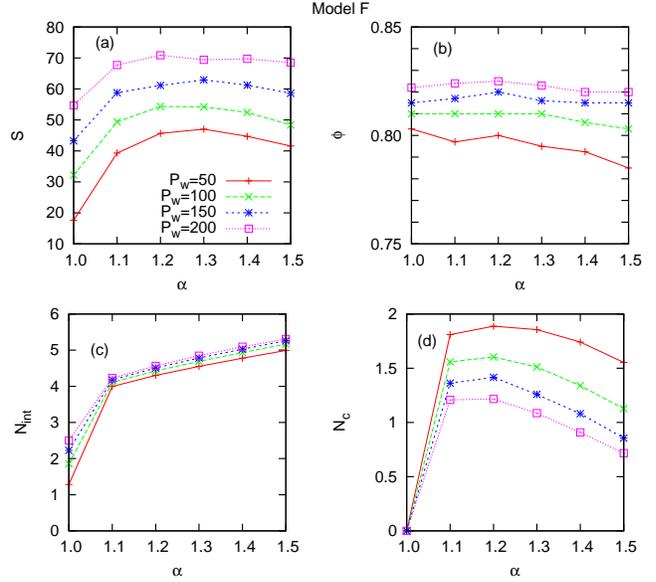}
\caption{
The interaction range $\alpha$ dependence of 
the shear stress $S$ (a), the packing fraction $\phi$ (b),
the number of bonds per particle $N_{int}$ (c),
and the number of cohesive bonds per particle $N_{c}$ (d)
in Model F for various pressure $P_w$.
$U=18$, which gives $\dot \gamma\approx 1$.
}
\label{variousP}
\end{figure}
First we check whether a plain shear flow can be realized 
in the present simulation.
Figure \ref{profile} (a) and (b) shows the velocity profile 
and the packing fraction 
for Model F with $U=18$, $P_w=100$, and
$\alpha=1.3$. 
From Fig.~\ref{profile}(a), we see that
the realized shear rate is almost constant with $\dot\gamma \approx 1$,
whereas some spatial structure can be seen in the packing fraction in 
 Fig.~\ref{profile}(b).
Similar profiles has been obtained
with various $\alpha$ and $P_w\ge 50$ for Model F.
However, we found that, when the wall pressure $P_w$ is too weak 
and/or the system size is too big,
most of the grain may stick to one of the walls
because of the cohesion,
and the velocity gradient is confined close to the other wall.
This tendency is stronger for Model P with large $\alpha$
and large enough system (Fig.\ref{profile}(c) solid line (L=40)).
When the system is small enough
(Fig. \ref{profile}(c) dashed line (L=20)),
the plain shear can be obtained for Model P also.
In the following we focus on the parameter range where
plain shear flow can be realized approximately.

\begin{figure}[hpt]
\includegraphics[angle=-90,width=0.4\textwidth]{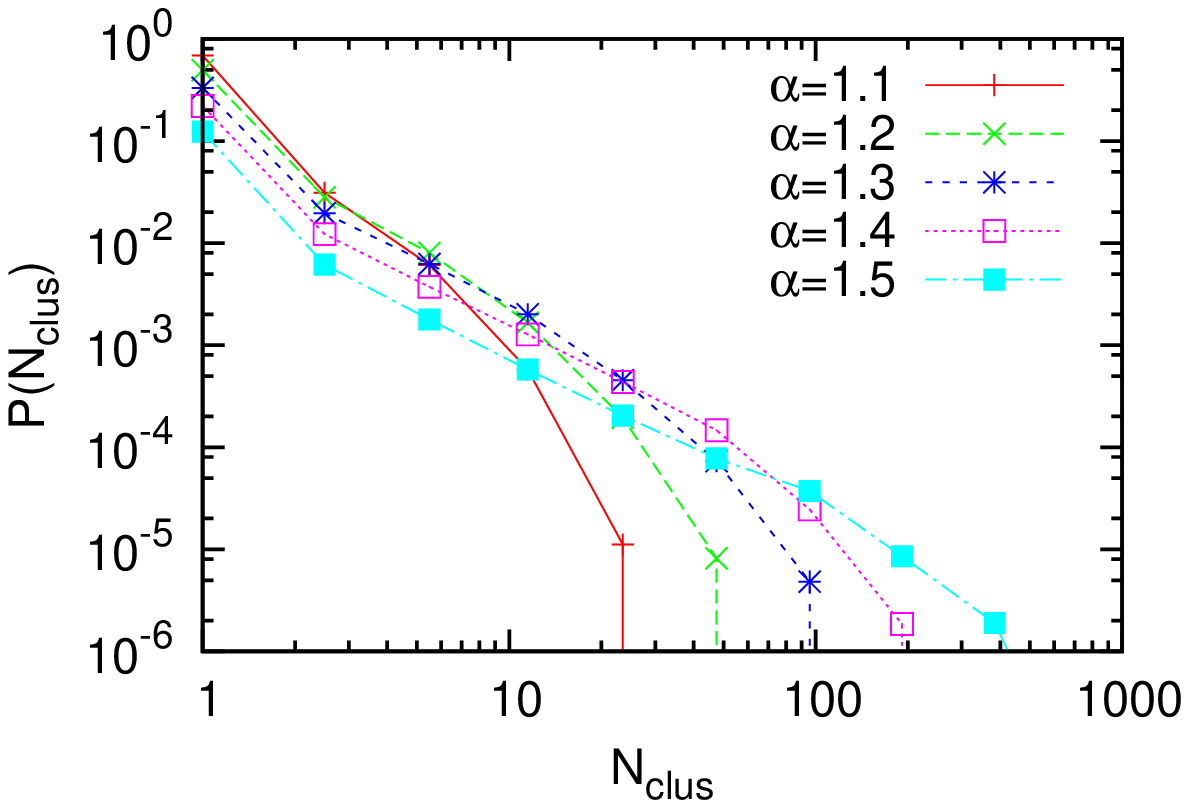}
\caption{
The cluster size distribution $P(N_{clus})$ with
$P_w=100$ and $\dot \gamma \approx 1$ 
for various interaction range $\alpha$.
}
\label{cluster}
\end{figure}

The interaction range $\alpha$ dependence of the 
shear stress $S$ is shown in Figure \ref{shearstress}
in the case of $P_w=100$ for Model F and Model P
with $\dot \gamma\approx 1$.
(for Model F, $U=18$ is used, while 
for Model P the small system with $L=20$ and
$N=400$ with $U=9$ is studied to avoid the non-uniform shear).
Note that, for $\alpha=1$ (no cohesion due to liquid bridge), 
both models should give the same results.
In Model P, the shear stress increases with $\alpha$, 
because larger $\alpha$ simply results in more cohesive interactions.
In Model F, on the other hand, $S$ shows a maximum around 
$\alpha=1.2$.

The maximum of the shear stress $S$ 
as a function of $\alpha$ in Model F can also be seen
for various values of pressure $P_w$, but the location of the 
maximum changes as can be seen in Fig.~\ref{variousP}(a).
The packing fraction $\phi$ at the 
middle of the system, 
is plotted against $\alpha$ for various $P_w$
in Fig.~\ref{variousP}(b); 
The packing fraction is higher for larger $P_w$
as expected, but the dependence on $\alpha$ is rather weak.

The existence of the maximum in the shear stress should be 
due to the competition of the increasing interacting particles
and the liquid-cluster formation which reduces the cohesive interaction. 
To see this more directly, the average number of interacting particles 
(``bonds'') per particle $N_{int}$
and the number of cohesive bonds 
(i.e., the bond that satisfies $N_{i,j}<n_{th}$ and $\Delta<0$)
per particle $N_c$
are shown in Fig.~\ref{variousP}(c) and (d), respectively.
We can see that, even though the interacting particles
increase with $\alpha$, the cohesive bonds has a maximum
as a function of $\alpha$. The location of the maxima are found to 
be slightly different between the shear stress $S$
and the number of cohesive interactions $N_{c}$, 
but they have a similar trend that the maximum moves 
towards smaller $\alpha$ for larger $P_w$.
We also found that $N_{c}$ is actually 
larger for smaller $P_w$ 
although $N_{int}$ is larger for larger $P_w$. 
This can be understood as an effect of the packing fraction $\phi$, 
i.e., for larger $P_w$, 
the packing fraction $\phi$ increases (Fig.~\ref{variousP}b),
which results in larger $N_{int}$
but $N_c$ decreases because
many of the cohesive interactions become inactive as
$N_{i,j}$ tends to exceed $n_{th}$ more easily.

The decrease of the cohesive bond upon increasing $\alpha$
should be a direct result of the increasing size of liquid clusters.
We characterize the liquid cluster size by
$N_{clus}$,  
the number of particles connected by bonds with $N_{i,j}\ge n_{th}$.
The cluster size distribution $P(N_{clus})$ is shown in Fig.~\ref{cluster}
for $P_w=100$ and $\dot \gamma\approx 1$.
We find that the distribution show exponential 
decay for small $\alpha$, but 
the decay becomes slower as $\alpha$ increase, 
and shows power-low like behavior 
at around $\alpha\approx 1.4 $, beyond which 
we get a finite probability to 
find a cluster as big as the system size.
The cluster size distribution depends on 
the pressure $P_w$ and the shear rate $\dot \gamma$,
but the tendency to have an exponential decay for small $\alpha$ 
and the slower decay for larger $\alpha$ is seen for all the
cases (data not shown).

\section{Summary and Discussion}
We proposed a simple model for the wet granular media
in the funicular state.
We assumed that 
the cohesive force acts between the particles connected by the 
liquid-gas interfaces but not those in a liquid cluster.
We examined the model behavior upon changing the liquid content,
which is represented by the interaction range parameter $\alpha$.
The shear stress $S$ in a steady shear flow 
was demonstrated to show a maximum at a certain value of $\alpha$.
The existence of the shear stress maximum is 
a result of competition between the two effects upon 
increasing the liquid content, i.e.,
the increase of interacting bonds, which increases the 
shear stress, and the formation of liquid clusters,
which reduces the cohesive interaction, thus reduces the shear stress.

In experiments, 
the shear modulus has been demonstrated to show a maximum 
upon increasing the liquid content \cite{MB07}.
The measurements have been done for small enough stain at finite
frequencies using rheometer.  The observed shear modulus is fairly
independent of the frequency in the range between 0.01 Hz and 10 Hz,
and found to show maximum values around the transition regime between
the pendular and the funicular regimes for various granular materials.
This experimentally observed maximum in the shear modulus may
be interpreted as a result of the competition between the formation of
liquid bonds and liquid clusters discussed above.

Another example of experimental observation 
is on the rotating drum\cite{XOK07}.  
Within the pendular state, it has been shown that the surface angle 
increases as the liquid volume increases at the low rotation rate, 
while at the high rotation rate the angle decreases 
with the liquid volume.  The behavior at the low rotation 
rate is consistent with that of our Model P, but
that at the high rotation rate is not.
The latter case seems to indicate that the lubrication and viscous
forces are important at higher rotation rate, and the rheological
properties may not be simulated by simple two-body cohesions.

Before concluding, let us discuss limitations and 
possible extensions of our model.

Firstly, the lubrication effect is not included in 
the present model, and that will be important for 
high shear rate cases as mentioned above. 

Secondly, the present model should not be valid for both limiting cases of the
small liquid content and the fully saturated regime.  In the former
case, the friction 
should become important 
for both the sliding \cite{RRWNC06} and
the rolling \cite{Bartels05,Gilabert07,Gilabert08} 
motions
because the cohesive force
pulls the particles together 
to contact even without external pressure 
while 
the liquid lubrication suppresses the 
friction effect
for larger liquid content. 
In 
the fully saturated regime, the rheology will be determined by the
hydrodynamic interaction, which is not included in the present model.

Lastly, the liquid content is only represented by the parameter $\alpha$, and
its conservation is not taken into account even though this may affect the
configuration of the interaction network. The model for the pendular
state proposed by Richefeu et al. \cite{RYR06, RRY06} 
is one of the possible approach
to include the liquid conservation effectively, where a liquid volume
dependent interaction has been adopted and the liquid is redistributed
to each liquid bridge according to a simple rule along the time
evolution.  This model showed the saturation of cohesion effects as a
function of the liquid content. Thus, if one add the liquid cluster
criterion to it as in our model, the drop of the cohesion effect at high
liquid content will be also observed because the liquid cluster will
reduce the cohesive bonds.  It will be interesting to see how the
behavior of the model changes by including these effects.

\section{Acknowledgment}
This work is partially supported by the grant-in-aid (21540418) 
by Japan Society for the Promotion of Science (JSPS).

\bibliography{WetGranular,bussei}
\bibliographystyle{eplbib}

\end{document}